\documentclass[letterpaper]{article} % DO NOT CHANGE THIS
\usepackage[]{aaai2026}  % DO NOT CHANGE THIS % submission
\usepackage{times}  % DO NOT CHANGE THIS
\usepackage{helvet}  % DO NOT CHANGE THIS
\usepackage{courier}  % DO NOT CHANGE THIS
\usepackage[hyphens]{url}  % DO NOT CHANGE THIS
\usepackage{graphicx} % DO NOT CHANGE THIS
\usepackage{natbib}  % DO NOT CHANGE THIS AND DO NOT ADD ANY OPTIONS TO IT
\usepackage{caption} % DO NOT CHANGE THIS AND DO NOT ADD ANY OPTIONS TO IT
\usepackage{algorithm}
\usepackage{algorithmic}

\usepackage{booktabs}

\usepackage{newfloat}
\usepackage{listings}
\DeclareCaptionStyle{ruled}{labelfont=normalfont,labelsep=colon,strut=off} % DO NOT CHANGE THIS
\floatstyle{ruled}
\newfloat{listing}{tb}{lst}{}
\floatname{listing}{Listing}

\usepackage{hyperref}
\nocopyright

\title{AI Assistance for Human Review of Default Judgments}
\author{
    Theodora Worledge\textsuperscript{\rm 1}\thanks{Corresponding author: worledge@stanford.edu},
    Othman Bensouda Koraichi\textsuperscript{\rm 2},
    Daniel Bernal\textsuperscript{\rm 2},
    Aviv Caspi\textsuperscript{\rm 2,3},\\
    Tatsunori Hashimoto\textsuperscript{\rm 1},
    Carlos Guestrin\textsuperscript{\rm 1},
    David Freeman Engstrom\textsuperscript{\rm 2,4}
}
\affiliations{
    \textsuperscript{\rm 1}Stanford Computer Science, Stanford University\\
    \textsuperscript{\rm 2}Deborah L. Rhode Center, Stanford Law School\\
    \textsuperscript{\rm 3}University of Chicago Law School\\
    \textsuperscript{\rm 4}Stanford Law School\\
}

\begin{document}

\maketitle
\pagestyle{plain}
\thispagestyle{plain}

\begin{abstract}
Overwhelmed courts in the United States review millions of default judgments each year. Unfortunately, such manual reviews are time-consuming and prone to error. In an audit of 188 debt collection cases granted default judgment by the Superior Court of Los Angeles, we find that 4\% contained major defects that should have entirely prevented default judgment, 10\% contained inconsistencies requiring reduced judgments, and 32\% contained errors requiring amendment prior to judgment. To support courthouses in default judgment review, we collaborated with courthouse attorneys and judges in designing a Default Assistant. The Default Assistant employs large language models to evaluate a case with respect to predetermined legal requirements and provide cited recommendations for an expert user's review. We equip users to verify these recommendations by grounding the assistant's explanations in cited quotes and tables from the original case filings. We conduct a controlled study with 66 law students that conservatively simulates court review, with more time and resources than court staff. We nevertheless find users aided by the Default Assistant were 6.0\% more accurate on the average requirement than unaided reviewers (p $<$ 1.0e-4). Simultaneously, users were 25.9\% faster in reviewing the average requirement than unaided reviewers (p $<$ 2.5e-10). Statutory requirements demanding extensive document search realized the largest gains, with error reductions and time savings from AI assistance up to 62\% and 34\%, respectively, relative to unassisted user performance and with differences statistically significant (p $<$ 0.05). Our work provides a proof-of-concept that AI assistants with citations have the potential to help resource-constrained courts conduct default judgment review more accurately and efficiently. 
\end{abstract}

% Uncomment to add links (ensure they do not de-anonymize you):
% \begin{links}
%     \link{Code}{https://aaai.org/example/code}
%     \link{Datasets}{https://aaai.org/example/datasets}
%     \link{Extended version}{https://aaai.org/example/extended-version}
% \end{links}

\section{Introduction}

The American civil justice system is failing to meet the needs of millions of litigants. Many of the 15 million civil cases filed in American state courts each year represent personal crises—a debt collection that results in wage garnishment, an eviction that leads to homelessness \citep{johnson2023one, garnham2022new}—that fuel cycles of unemployment, poverty, poor health, and family breakdown \citep{mullen2019fifty, desmond2015eviction}. Yet, despite these high stakes, roughly three-quarters of these civil cases involve at least one person who cannot afford a lawyer \citep{agor2015landscape}. Many defendants do not take action to defend themselves in court; defendants only respond in 6\% of debt collection cases filed at the Superior Court of Los Angeles County (SCLAC) \citep{johnson2023one}. With minimal adversarial process to surface evidence of defects in cases, courts, constrained by the impracticality of rigorous manual review, are more likely to issue erroneous default judgments \citep{jimenez2015dirty, bookman2024default}.

The court caseload for default judgment review is crushing. Each year, SCLAC—the largest trial court in the nation\footnote{https://www.lacourt.ca.gov/pages/lp/court-communications/tp/about-the-court}—routes as many as 30,000 debt collection default judgment requests to court staff for manual review. The high caseload creates severe time constraints. SCLAC research attorneys spend on average four minutes per case to verify over a dozen statutory requirements.\footnote{Estimated from the number of cases requesting a default judgment per week at SCLAC and the number of dedicated attorney-hours per week.} The failure of any requirement influences the judge's decision to grant default judgment in favor of debt buyers pursuant to CA Civil Code §§ 1788.58-60. Given such resource constraints, even the most diligent courts are prone to error. 

Using annotation guidelines for statutory and procedural requirements developed by legal academics and court professionals, we audited 188 debt-buyer plaintiff collections cases granted default judgment by SCLAC. We found that 4\% contained major defects that should have entirely prevented default judgment, such as a complaint that violated the statute of limitations or a case missing essential documentary evidence. 10\% of the cases contained inconsistencies that should have resulted in a reduction in the amount of judgment requested by the plaintiff. Moreover, 32\% contained errors that should have resulted in amended petitions prior to judgment. These estimates indicate there may be over a thousand erroneous debt collection default judgments per year in Los Angeles alone.

Courts across the United States need support in efficiently identifying unsatisfied statutory requirements in default judgment review.  Although hiring more human reviewers would help, the number of collections cases is rising at SCLAC (Appendix Figure \ref{fig:collection_filings}) and more broadly across California \citep{johnson2023one}. Also, the number of collections filings may continue to rise as debt buyers leverage artificial intelligence (AI) to file at higher rates \citep{shah2026access}. Courts should consider assistive AI to support expert human review. We believe that an assistive tool has the potential to significantly reduce the rate of unjust case outcomes and the life-altering consequences that follow.

Furthermore, courts are currently too capacity-constrained to reexamine their role in ensuring accurate outcomes. An assistive tool can help give courts the necessary capacity to rethink the policies they implement and enforce. Indeed, several judges at SCLAC are even now considering interpreting hearsay requirements more restrictively, despite higher costs in manual review. 

Through an interdisciplinary collaboration with legal academics, SCLAC judges, SCLAC research attorneys, and computer scientists, we propose a ``Default Assistant'' that provides requirement-level recommendations to court staff and flags cases that are more likely than others to contain errors. The Default Assistant is built using large language models (LLMs) to efficiently parse pages of text and tables to identify evidence satisfying or failing statutory requirements. Importantly, we design the assistant to ground its recommendations in the case filings and cite the relevant quotes and tables in each accompanying explanation. The Default Assistant will support courts in upholding statutory requirements by providing recommendations, surfacing relevant evidence through precise citations, and leaving final decisions up to staff attorneys and judges.

In this work, we evaluate an initial version of the Default Assistant with law students on default judgment requests filed by debt-buyer plaintiffs. We find experimental evidence that the Default Assistant can improve case review: assisted users were 6.0\% more accurate and 25.9\% faster than unassisted users. Towards the end of supporting courts in issuing accurate and timely default judgments, our work makes the following contributions:

\begin{enumerate}
\item{\textbf{Task Definition and Dataset Generation:} Our team of legal academics collaborated with SCLAC experts to define detailed case annotation guidelines for relevant statutory and procedural requirements. We work with research assistants to annotate defects of 188 California debt collection cases granted default judgment.}

\item{\textbf{Default Assistant Implementation:} We design and develop an LLM-based Default Assistant in accordance with the previously identified California statutory and procedural requirements, working closely with court stakeholders.}

\item{\textbf{Human Evaluation:} We compare the performance of humans using Default Assistant recommendations to that of unassisted humans. We find that the assisted humans are faster and more accurate at debt collection default judgment review.}

\item{\textbf{Fairness Evaluation:} Using augmented defendant names, we find no meaningful differences in standalone Default Assistant accuracy across defendants’ race and gender.}
\end{enumerate}  

\section{Related Work}

% XAI for human-AI teamwork 
\textbf{Designing for Human-AI Collaboration.} Given the necessity of human oversight in the judicial setting, a critical feature of the Default Assistant is to equip a human user to verify recommendations. Prior works propose different forms of explainable AI to enhance human-AI teamwork. Local explainability methods, such as LIME \citep{ribeiro2016model}, provide insight into which parts of an input were influential in the AI output. LLMs can provide natural language explanations of their answers. 

However, local and free-form explanations have been shown to increase over-reliance on incorrect model recommendations \citep{bansal2021does, jacobs2021machine, 10.1145/3706598.3714020}. Furthermore, \citet{turpin2023language} show that LLM-generated explanations are not necessarily faithful to the model's underlying reasoning. A promising line of work shows that explanations that enable users to verify model recommendations with relevant evidence, rather than those that hint at the reasoning process, can reduce over-reliance and improve user performance \citep{fok2024search, 10.1145/3706598.3714020}. Accordingly, we engineer a Default Assistant that provides recommendations with citations to case files to equip users in efficiently verifying the assistant and catching mistakes. 

% Citation methods (pull from methods section)
\textbf{Methods for Cited Generation.} Methods for generating cited outputs differ primarily in when citations are produced relative to the generated text. One line of work generates text and citations jointly, training or prompting models to produce inline citations alongside their answers \citep{nakano2021webgpt, menick2022teaching, shao-etal-2024-assisting, jiang-etal-2024-unknown, zhang-etal-2025-longcite}. Another generates first and attributes after the fact, revising the output with sources retrieved post hoc \citep{gao2022rarr, li-etal-2024-citation}. 

Prior work provides empirical support for a third design choice: \textit{attribute-first-then-generate}, in which supporting evidence is identified before generation rather than alongside or after it. \citet{slobodkin2024attributefirst} demonstrate that attributing evidence prior to generation yields fewer unsupported claims than the simultaneous generation of answers and citations. \citet{worledge2024entailed} further show that generating model outputs from pre-identified source quotes improves citation precision and coverage over deployed systems, including post hoc citation. We follow this three-step process of identifying, verifying, and finally generating from source quotes for precise and comprehensive citations, enabling transparent tracing of each model generation to specific lines of the underlying case materials.

% Other human-AI evals of legal systems (see Aileen's paper!)
\textbf{Evaluating Legal Human-AI Collaboration.}

Because the success of human-AI teams is influenced by factors beyond AI system performance \citep{vasconcelos2023explanations}, we run a randomized controlled trial to evaluate the performance of humans using the Default Assistant for debt collection default judgment review. Prior work has studied AI-assisted users on other legal tasks. \citet{choi2024lawyering} run a controlled study with 59 law students on legal drafting assignments, observing large efficiency gains from GPT-4 access with insignificant or small changes in work quality. \citet{https://doi.org/10.1111/jels.12396} compare AI summaries and AI highlighting of legal complaints in a controlled experiment with 206 law students, finding that highlighting reduces task time by 30\% without a measured quality drop, while summaries alone produce neither efficiency nor quality improvements. \citet{tan2025licensegpt} test a fine-tuned LLM for dataset license compliance with four software IP lawyers, finding substantial time savings without quantifying changes in accuracy. Similar to \citet{https://doi.org/10.1111/jels.12396}, we consider AI outputs with explainability features (i.e., highlights and citations) and like \citet{tan2025licensegpt}, we evaluate a custom AI system. Our work may be the first study of human-AI collaboration on a legal task to show significant and meaningful increases in both efficiency and accuracy.

\section{Socio-Technical Discussion}
Especially since the advent of LLMs, automated tooling has been considered and deployed for a litany of legal tasks, including legal research, case review, and contract analysis \citep{siino2025exploring, pasquale2019rule}.\footnote{The development of legal AI for law firms is moving a mile-a-minute; Harvey and Legora are two prominent companies in this space.} In response to the possibility of assistive AI in law and other high-stakes settings, prior literature has characterized the risks of such deployments. 

In this section, we discuss how the design of our Default Assistant and controlled experiment are informed by our multidisciplinary team and SCLAC stakeholders in light of these risks. Our contributors include a law professor of legal ethics, a legal expert in debt collection, AI researchers, and SCLAC judges, leadership, and court research attorneys. We have identified three areas of risk: automation bias, historical bias, and technological lock-in. 

A primary challenge of assistive AI is automation bias: the tendency for users to blindly rely on AI recommendations \citep{ruschemeier2024automation}, especially due to opaque model reasoning that hinders fact-finding and verifiability \citep{kuzniacki2022towards, koenecke2025tasks}. Automation bias is most damaging when it leads users to accept incorrect AI recommendations on tasks they would have completed correctly on their own. To mitigate this risk, we design the Default Assistant to cite each recommendation to specific lines in the case materials, aligning with broader legal norms that use citations to indicate source provenance. We also center our evaluation around a controlled user study, rather than a standalone system evaluation, and conduct an analysis on over-reliance and appropriate reliance. 

Second, prior works highlight the risk of allocational disparities arising from deployed AI that amplify historical bias, along identity-based attributes \citep{ajunwa2019paradox, angwin2022machine}. In the development of our Default Assistant, it is possible that biases are learned from the outcomes in our debt collection default judgment dataset or from the training data of the underlying LLM itself. To mitigate the amplification of these biases, we annotate debt collection court cases with new ground-truth labels, rather than using the historical SCLAC decisions, which may be unduly constrained by time and resource limitations. To detect signs of bias in our end-to-end system, we also conduct an evaluation of the stability of Default Assistant recommendations across different defendant names perceived to be of different race and gender identities.

Lastly, \citet{crootof2019cyborg} explains that reliance on software with high technical overhead can add friction to legal evolution in courts. While this work takes substantive steps to address automation bias and historical bias, technological lock-in remains an open problem. We further discuss this concern in our limitations section. While the technical framing of our work is no panacea for these multi-faceted challenges—which also require broader AI governance and institutional policies—these concerns have meaningfully shaped the design and evaluation of our system.

\begin{figure*}[h]
  \centering
  \includegraphics[width=1\linewidth]{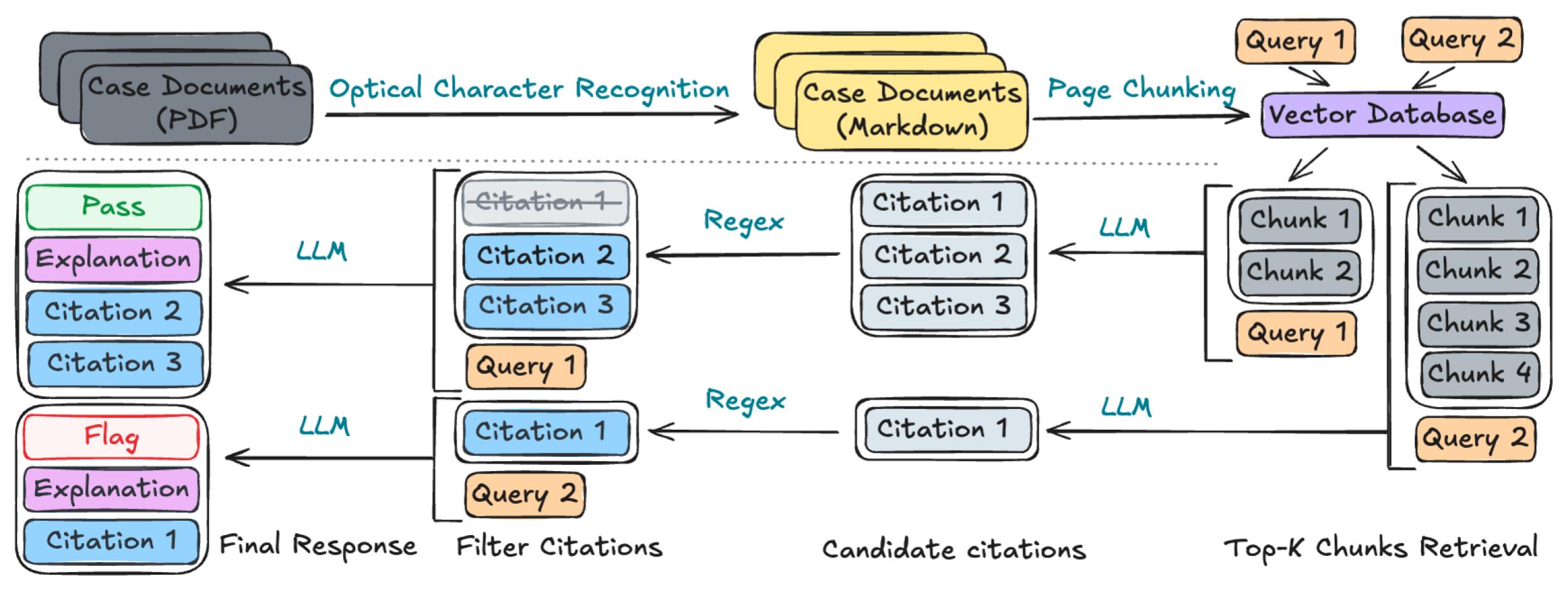}
  \caption{The Default Assistant processes PDF case documents to provide cited recommendations for case requirements.}
  \label{fig:pipeline}
\end{figure*}
\section{Data}

We collected an initial sample of 200 debt collection cases filed in 2023-2024 by debt buyer plaintiffs. Specifically, we sampled 100 cases that requested default judgments and 100 cases that were granted default judgments. We removed two cases filed by original creditor plaintiffs (mistakenly labeled as debt-buyer cases in the court's filing system), two cases that were rejected prior to default judgment review due to administrative errors, and nine cases with fewer than three independent annotations from unique research assistants (methodology described below). Our final dataset is comprised of 96 cases that were sampled to have a request for default judgment and 92 cases that were granted default judgment. We note that the 92 cases sampled to have a request were all ultimately granted default judgment, resulting in a total of 188 cases that were all granted default judgment. We report historical court accuracy on these 188 cases. From this sample, we used 20 random cases for the development of the Default Assistant and held out 168 cases for a robust evaluation set.   

All of the debt collection case files are public records and were accessed via the electronic filing and case management systems at SCLAC. The PDF files for each case include a complaint, request for default judgment, and often one or more declarations containing business record evidence. On average, there are about 55 pages of PDF content per case. These files form the evidentiary basis for default judgment review, yet vary widely in format—some are text-based, while others are scanned images. 

Our research team, led by a practicing lawyer and supervised by a law professor, collaborated with court judges and research attorneys to develop annotation guidelines grounded directly in procedural and statutory requirements (CA Civil Code \S\S 1788.58–.60). The requirements and instructions for conducting the review of each requirement on an individual case were subject to multiple rounds of revision between our team and our court collaborators. Once finalized, the lawyer on our team trained law student research assistants to follow these guidelines to annotate sub-requirements, which we combined into higher-level \textit{requirements}, using the logical \texttt{AND} operator. For each annotation, assistants marked sub-requirements as ``Satisfied'', ``Not Satisfied'', or ``Unclear''. Only 11 sub-requirement annotations out of 8,400 were marked ``Unclear''.\footnote{When evaluating the Default Assistant, we convert ``Unclear'' annotations to ``Not Satisfied'' because the assistant should flag any case that requires further scrutiny. However, we do not count ``Unclear'' annotations toward major or minor error categories in our reported historical case outcomes; we instead provide lower bounds on the number of errors in the sample.} On average, the research assistants spent 14.1 minutes applying the guidelines to a case. 

\begin{table}[h!]
\caption{Inter-annotator agreement and Cohen's $\kappa$ for the annotations, prior to two-thirds majority aggregation into ``gold labels''. The measures of Cohen's $\kappa$ indicate substantial agreement (0.61–0.80) in three requirements and moderate (0.41–0.60) in two (n=188). Annotators only classified cases as valid for the \textit{Agreement to Debt} and \textit{Sworn Declaration} requirements. In this table, ``Unclear'' annotations are counted as ``Not Satisfied''.}
\label{tab:inter_annotator_kappa}
\centering
\begin{tabular}{lcc}
\toprule
Requirement & Agreement & Cohen's $\kappa$ \\
\midrule
Debtor Address & 0.95 & 0.77 \\
Chain of Title & 0.97 & 0.65 \\
Request-Complaint Cons. & 0.90 & 0.61 \\
Last Payment Date \& SoL & 0.88 & 0.60 \\
Charge-Off Balance & 0.92 & 0.50 \\
Agreement to Debt & 1.00 & - \\
Sworn Declaration & 1.00 & - \\
\bottomrule
\end{tabular}
\end{table}

The Cohen’s kappa coefficients \citep{cohen1960coefficient} in Table \ref{tab:inter_annotator_kappa} show that our annotation guidelines and procedure achieve substantial to moderate inter-annotator agreement across all requirements \citep{hall2008systematic}. These agreement rates could be improved, so we strengthened our labels by collecting a third round of independent annotations and take the two-thirds majority label for each requirement over the three independent annotations from different research assistants for each case. We refer to these strengthened annotations as the ``gold labels''. These gold labels are used in the evaluation of both the historical court performance and the Default Assistant.   

\subsection{Statutory and Procedural Requirements}
\label{subsec:stat_reqs}

Working closely with court stakeholders, our legal experts identified the list of requirements that the court reviews for default judgment requests in debt-buyer plaintiff collections cases, as of September 2025. The gold coding methodology and the Default Assistant both evaluate each requirement listed below through a decomposition of sub-requirements, presented in Appendix Figure \ref{fig:default_assistant_graph}.\footnote{Requirement-level recommendations from the assistant are obtained by the same procedure used to obtain requirement-level gold labels: if a case fails any sub-requirement, it fails the parent requirement.} The seven legal requirements we study are as follows:

\textit{Request-Complaint Consistency}: The Request for Default Judgment must not exceed the complaint’s prayer in damages or interest and may only request attorney fees or costs of suit if those were included in the complaint's prayer.

\textit{Agreement to Debt}: The complaint must include a copy of a contract or a monthly credit statement recording a purchase transaction, last payment, or balance transfer to show the defendant’s agreement to the debt (§ 1788.60 (b)).

\textit{Sworn Declaration}: The case files must include at least one declaration signed under penalty of perjury in support of default judgment, signed by an individual with personal knowledge of the relevant business records (§ 1788.60 (a)).

\textit{Charge-Off Balance}: The complaint must allege the charge-off balance (§ 1788.58 (a)(4)) and evidence in a sworn declaration must substantiate the alleged amount (§ 1788.60 (a)).

\textit{Last Payment Date \& Statute of Limitations (SoL)}: The complaint must allege the date of last payment (§ 1788.58 (a)(5)), evidence in a sworn declaration must substantiate the alleged date (§ 1788.60 (a)), and the complaint filing date must fall within the four-year statute of limitations from a substantiated date of last payment. 

\textit{Debtor Address}: The complaint must allege the name and last-known address of the debtor from the charge-off creditor (§ 1788.58 (a)(7)) and evidence in a sworn declaration must substantiate the alleged address (§ 1788.60 (a)).

\textit{Chain of Title}: The complaint must allege the names and addresses of all post-charge-off purchasers of the debt (§ 1788.58 (a)(8)) and evidence in a sworn declaration must substantiate the alleged chain of ownership (§ 1788.60 (a)).

We further categorize these requirements into major and minor error designations. \textit{Major - Reject} errors are severe case defects that would preclude default judgment (i.e., time-barred cases, failure to include a valid declaration, and missing substantiation of the chain of title). \textit{Major - Reduce} errors are inconsistencies––between the amounts requested in the complaint and default judgment––that would reduce the amount of the judgment. Finally, \textit{Minor - Amend} errors are all of the remaining requirements–technical errors that would likely be remedied by an amended complaint. 

\section{Default Assistant Implementation}

\begin{figure*}[h]
    \centering
    \includegraphics[width=\linewidth]{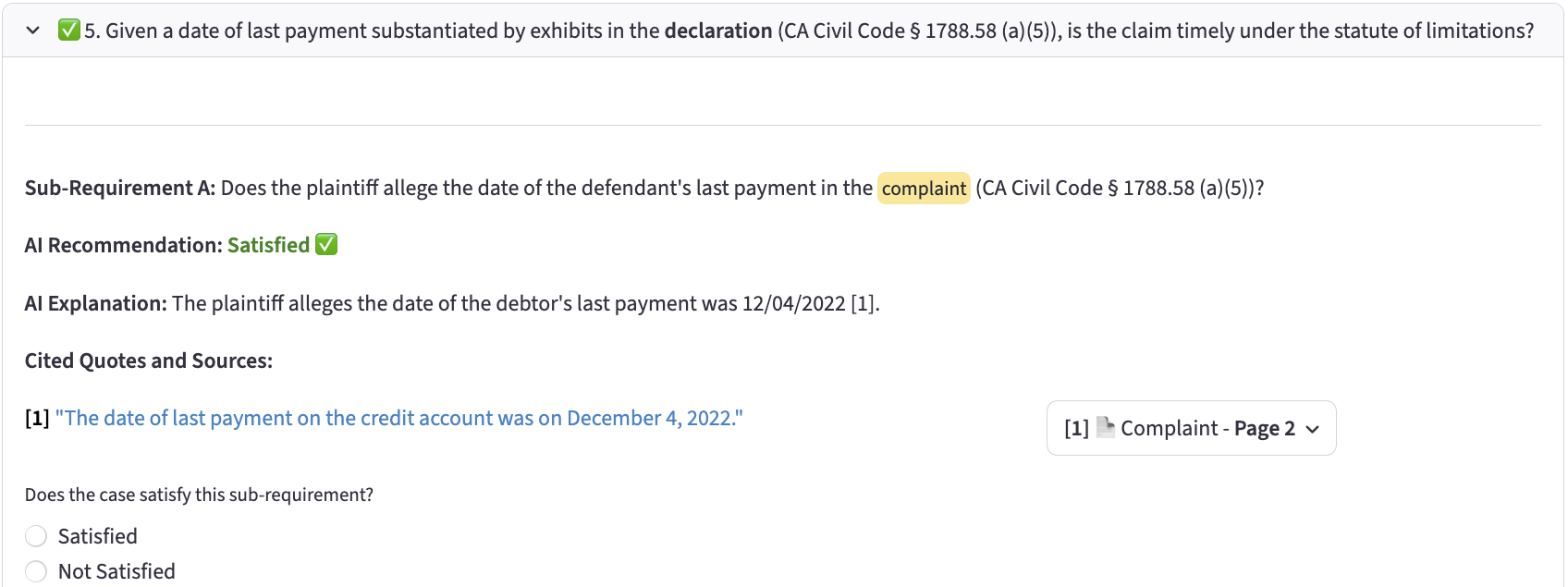}
    \caption{For each recommendation, the Default Assistant provides a binary recommendation and a free-form explanation cited to verified quotes from the case materials. Users may also view the original case file PDF page as shown in Appendix Figure \ref{fig:ui_subreq_expanded}.}
    \label{fig:citation_example}
\end{figure*}

We built an AI ``Default Assistant'' to flag case defects and provide recommendations for compliance with CA Civil Code \S\S 1788.58–.60 and other procedural requirements. If the Default Assistant finds inadequate evidence for a requirement, the assistant flags the requirement for court staff, who can verify the recommendation and make the ultimate decision. To facilitate verification, the Default Assistant includes citations to case materials for each recommendation, such as short quotes or tables that are highlighted within the original PDF (Figure \ref{fig:citation_example} and Appendix Figure \ref{fig:ui_subreq_expanded}).

The Default Assistant processes each case file through a structured pipeline that converts PDFs into Markdown using optical character recognition (OCR), and chunks and stores the files by page in a vector database. For each sub-requirement, the assistant performs retrieval-augmented generation (RAG) with additional citation steps to generate a recommendation (Figure \ref{fig:pipeline}). The model grounds each recommendation in verified evidence from the original filings and provides an explanation citing specific quotes and tables. We define each round of retrieval and cited generation as a node in LangGraph\footnote{www.langchain.com/langgraph}—a graph that asynchronously executes nodes in parallel and in order of dependency, permitting later nodes to use information identified in previous nodes (Appendix Figure \ref{fig:default_assistant_graph}). For each node, we wrote a retrieval query and a generation query. We iteratively improved these queries by evaluating performance on the development set. We provide additional implementation details in the following subsections. 

\subsection{Data Processing and Semantic Retrieval}

After empirical testing and manual inspection of ten representative cases, we opted for Azure AI Document Intelligence as our OCR engine given multiple open and closed source options. The Azure service preserves Markdown layout cues such as tables, headers, and key-value pairs, enabling the preservation of structured elements in the retrieval and LLM cited generation pipeline. 

Following OCR, we chunk the Markdown text along the original page boundaries from the PDFs. Chunking along page boundaries ensures that tables, footers, and section headers remain coherent within the same chunk. Each page-level chunk is then embedded using OpenAI's \texttt{text-embedding-3-small} and stored in a vector database for semantic retrieval. Top-$k$ retrieval was set to return either 4 or 10 pages, depending on the requirement. Failures to retrieve relevant chunks were rare on the development set. 

\subsection{Cited Generation}

To ensure that all Default Assistant recommendations are grounded in verifiable evidence, the assistant employs an \textit{attribute-first-then-generate} citation framework \citep{slobodkin2024attributefirst}. For each step of case review, the Default Assistant first retrieves relevant documents from the vector database. Then, the LLM (\texttt{gpt-4.1}) is few-shot prompted to extract quotes and tables from the retrieved documents that are necessary to evaluate the current legal requirement. These extracted spans are then relocated in the original document; quotes are identified using fuzzy, white-space-agnostic matching, while tables are located using regex on the values. Each candidate citation text is replaced with its directly copied counterpart from the original case files. Only successfully located quotes and tables are retained and passed to the LLM with a few-shot generation query to compose the final response. The final response includes a binary recommendation that the legal requirement is ``Satisfied'' or ``Not Satisfied'' and a free-form explanation with citations to the verified source quotes. 

\section{Human Study Methodology}
\label{subsec:human_evaluation_methodology}

We recruited 66 participants to review debt collection default judgment requests. All participants were law students. There is reason to believe that law students are a reasonable study population; law students commonly graduate into clerkships and SCLAC is considering delegating debt collection default judgment review to clerks. Prior works studying AI assistance in legal workflows similarly work with law student participants for tasks relevant to law clerks and junior associates \citep{https://doi.org/10.1111/jels.12396, choi2024lawyering}. 

The study was four hours long and included training and case review to simulate court staff workflows. We co-designed an in-person, 80-minute training reflecting standard procedure with an SCLAC research attorney who specializes in default judgment review. The training covered relevant procedural and statutory requirements, illustrated by real case-file examples. The training also included practice reviews of acceptable and unacceptable debt-buyer cases, audience Q\&A, and printed reference materials for use during the study. Participants were randomly assigned to two conditions:
\begin{enumerate}
\item \textbf{Human:} No access to the Default Assistant
\item \textbf{Team:} Access to the Default Assistant
 \end{enumerate}
 
After training, participants were directed to separate rooms by condition and given branch-specific instructions and websites, with participants blinded to other conditions. Both the human and team websites displayed the same requirements, case information, links to case files, and a 10-minute countdown timer (Appendix Figure \ref{fig:ui_req}). Branches differed only in whether Default Assistant recommendations and citations were shown on the website.

During the simulation, participants reviewed 12 randomly assigned cases, deciding whether each requirement was “Satisfied” or “Not Satisfied.” Responses were saved at a 10-minute per-case limit, with any incomplete requirements automatically marked as satisfied, though participants could submit earlier. While court staff typically spend about four minutes per case, we allowed for more time to account for participant inexperience, setting the limit at 10 minutes based on pilot study averages.

The websites collected sub-requirement decisions, time spent per requirement, and self-reported user confidence scores on a 0-100 scale for each requirement. Some of the 168 cases in the held-out test set were randomly selected to receive decisions from multiple participants within a branch, resulting in 204 evaluations per branch. Participants were compensated with \$100 gift cards and free dinners. We consulted with the IRB prior to recruiting participants; the board confirmed that formal review and approval were not required. 

As with the gold labels and AI recommendations, we aggregate the sub-requirement decisions into requirement-level decisions. We refer to decisions from the human and team participants as \textit{accepting} or \textit{rejecting} and decisions from the Default Assistant as \textit{flagging} or \textit{not flagging} a case or requirement.  

\section{Human Study Results}

The Default Assistant (DA) helped users improve their average requirement accuracy per case by 6.0\%, while decreasing the average time spent reviewing each requirement per case by over 25\% (Table \ref{tab:overallRequirementResults}). Differences in average requirement accuracy and review time per case between humans with and without the Default Assistant were found to be statistically significant (p $<$ 0.002). Any increase in user confidence for those using the Default Assistant appears modest and tracks the increase in accuracy, however, the difference is not statistically significant.

\begin{table}[h!]
\caption{Users aided by the Default Assistant were both more accurate and faster than unaided users. This analysis considers the average paired accuracy, time, and user confidence (out of 100) over all requirements for a case (n=204). Starred values (*) indicate statistically significant non-zero differences after Bonferroni correction (p $<$ 0.002).}
\label{tab:overallRequirementResults}
\centering
\begin{tabular}{llll}
\toprule
Metric & Absolute Gain & Relative Gain & p-value \\
\midrule
Accuracy & 5.3 pp & 6.0\% & 1.0e-04* \\
Timing & -12.8s & -25.9\% & 2.5e-10* \\
Confidence & 2.2 & 3.0\% & 1.7e-01 \\
\bottomrule
\end{tabular}
\end{table}

Table \ref{tab:team_vs_human_results} shows the rates of error reduction and time savings by requirement in the team setting, relative to the human baseline. The Default Assistant helped users achieve 62\% fewer errors in judging the \textit{Charge-off Balance} requirement, 47\% fewer for \textit{Last Payment Date \& SoL}, and 48\% fewer for \textit{Debtor Address}, with differences significant (p $<$ 0.05). The Default Assistant also expedites review time by 24\% to 34\%, with the differences in time between the human baseline and the assisted human found to be significant (p $<$ 0.05) for all requirements but \textit{Agreement to Debt} and \textit{Chain of Title}.

\begin{table*}[h!]
\caption{Paired human and citation-assisted team performances by requirement, ordered by absolute accuracy gain. We report accuracy, relative error reduction, and relative time savings, taken with respect to the human baseline (n=204). Starred comparisons (**) indicate statistically significant non-zero differences (p $<$ 0.05) after Benjamini-Hochberg correction. Calculated at full precision; rounded for display.)}
\label{tab:team_vs_human_results}
\centering
\begin{tabular}{lccccccc}
\toprule
Requirement & Human Acc. & Team Acc. & Rel. Error Reduction & Rel. Time Savings \\
\midrule
Last Payment Date \& SoL & 0.78 & 0.88 & 47\%** & 34\%** \\
Charge-Off Balance & 0.86 & 0.95 & 62\%** & 30\%** \\
Debtor Address & 0.85 & 0.92 & 48\%** & 31\%** \\
Chain of Title & 0.88 & 0.93 & 44\% & 18\% \\
Agreement to Debt & 0.95 & 0.98 & 60\% & 7\% \\
Sworn Declaration & 0.97 & 0.99 & 57\% & 30\%** \\
Request-Complaint Consistency & 0.90 & 0.91 & 5\% & 24\%** \\
\bottomrule
\end{tabular}
\end{table*}

The Default Assistant enabled participants to achieve error reduction through increases in precision and, to a lesser extent, recall in rejecting defective cases (Table \ref{tab:true_acceptance_true_rejection}). The small number of defects limits interpretation; for instance, the apparent changes in recall for \textit{Last Payment Date \& SoL}, \textit{Debtor Address}, and \textit{Request-Complaint Consistency} correspond to differences over only one or two cases. However, we observe that human-AI teams generally increase precision over unassisted humans without evidence of reducing recall, and even appear to increase recall for \textit{Request-Complaint Consistency}. 

\begin{table*}[h!]
\caption{Precision and recall for identifying invalid cases by requirement for human and team decisions, ordered by number of defects (n=204). Gains are the team metrics minus human metrics. AI assistance increases or maintains precision for all requirements and increases or maintains recall of 6 out of 7 requirements. However, the low number of defects limits the interpretation of these measurements. Calculated at full precision; rounded for display.}
\centering
\label{tab:true_acceptance_true_rejection}
\begin{tabular}{lccccccc}
\toprule
\multicolumn{2}{c}{} & \multicolumn{3}{c}{Precision} & \multicolumn{3}{c}{Recall} \\
Requirement & Num. Defects & Human & Team & Gain & Human & Team & Gain \\
\midrule
Last Payment Date \& SoL & 32 & 0.41 & 0.57 & 0.17 & 0.91 & 0.97 & 0.06 \\
Debtor Address & 19 & 0.34 & 0.57 & 0.23 & 0.68 & 0.63 & -0.05 \\
Request-Complaint Consistency & 18 & 0.46 & 0.49 & 0.03 & 0.61 & 0.94 & 0.33 \\
Charge-Off Balance & 14 & 0.30 & 0.57 & 0.27 & 0.79 & 0.86 & 0.07 \\
Chain of Title & 6 & 0.12 & 0.21 & 0.09 & 0.50 & 0.50 & 0.00 \\
Agreement to Debt & 0 & 0.00 & 0.00 & 0.00 & 0.00 & 0.00 & 0.00 \\
Sworn Declaration & 0 & 0.00 & 0.00 & 0.00 & 0.00 & 0.00 & 0.00 \\
\bottomrule
\end{tabular}
\end{table*}

The Default Assistant improves accuracy across major and minor error categories between the unassisted human and team branches (Table \ref{tab:major_minor_performance}). \textit{Major - Reject} accuracy increased due to improvement in true rejections (+9 pp), i.e., the rejection of invalid cases, and improvement in true acceptances (+9 pp), i.e., the acceptance of valid cases. Gains in true rejections (+33 pp) drove improvements in \textit{Major - Reduce} accuracy, while gains in true acceptances (+13 pp) drove improvements in \textit{Minor - Amend} accuracy.  

\begin{table*}[h!]
\centering
\caption{The accuracy, rate at which invalid cases are rejected (True Rejection), rate at which valid cases are accepted (True Acceptance), and relative changes in true rejection/acceptance with AI assistance, taken with respect to the human baselines (n=204). \textit{Major - Reject}: case defects unlikely to be remedied by an amended complaint. \textit{Major - Reduce}: case defects that won't preclude judgment, but would reduce the amount of the judgment. \textit{Minor - Amend}: case defects that can be amended by the plaintiff without additional evidence. Calculated at full precision; rounded for display.}
\label{tab:major_minor_performance}
\begin{tabular}{lccccccccccc}
\toprule
 & \multicolumn{1}{c}{Num.} & \multicolumn{2}{c}{Accuracy} && \multicolumn{3}{c}{True Rejection} && \multicolumn{3}{c}{True Acceptance} \\
Requirement & Defects & Human & Team && Human & Team & Rel. Change && Human & Team & Rel. Change \\
\midrule
Major - Reject & 11 & 0.81 & 0.90 && 0.73 & 0.82 & 13\% && 0.82 & 0.91 & 11\% \\
Major - Reduce & 18 & 0.90 & 0.91 && 0.61 & 0.94 & 55\% && 0.93 & 0.91 & -2\% \\
Minor - Amend & 44 & 0.73 & 0.83 && 0.86 & 0.89 & 3\% && 0.69 & 0.82 & 19\% \\
\bottomrule
\end{tabular}
\end{table*}

\subsection{Complementary Performance}

\begin{table*}
\caption{Comparison of team ($T$), unassisted human ($H$), and Default Assistant ($DA$) accuracy across requirements, with highest (tied) accuracy per row in bold (n=204). Absolute gains and amplification ($\frac{T-H}{|DA-H|}$) quantify how collaboration often exceeds the DA–human gap. Ordered by team gain over DA. Calculated at full precision; rounded for display.}
\centering
\label{tab:complementary_performance}
\begin{tabular}{lcccccc}
\toprule
Requirement & $T$ & $H$ & $DA$ & $T-H$ & $DA-H$ & Amplification \\
\midrule
Last Payment Date \& SoL & \textbf{0.88} & 0.78 & 0.87 & 0.10 & 0.09 & 1.17x \\
Charge-Off Balance & \textbf{0.95} & 0.86 & 0.94 & 0.09 & 0.08 & 1.06x \\
Debtor Address & \textbf{0.92} & 0.85 & 0.89 & 0.07 & 0.04 & 1.88x \\
Chain of Title & \textbf{0.93} & 0.88 & 0.92 & 0.05 & 0.04 & 1.38x \\
Agreement to Debt & \textbf{0.98} & 0.95 & 0.96 & 0.03 & 0.01 & 3.00x \\
Sworn Declaration & \textbf{0.99} & 0.97 & 0.98 & 0.02 & 0.01 & 2.00x \\
Request-Complaint Consistency & 0.91 & 0.90 & \textbf{0.93} & 0.00 & 0.03 & 0.17x \\
\bottomrule
\end{tabular}
\end{table*}

The team setting generally achieves stronger performance than the unassisted humans and Default Assistant individually (Table \ref{tab:complementary_performance}). We observe that unassisted human performance falls behind the Default Assistant on every requirement; we present an amplification multiplier that indicates how much of the DA–human performance gap is actually realized—or counteracted—when humans collaborate with the Default Assistant. This multiplier indicates that the team setting compensates for and even exceeds the DA-Human performance gap for most requirements.

\subsection{Reliance}

\begin{figure*}[h]
  \centering
  \includegraphics[width=1\linewidth]{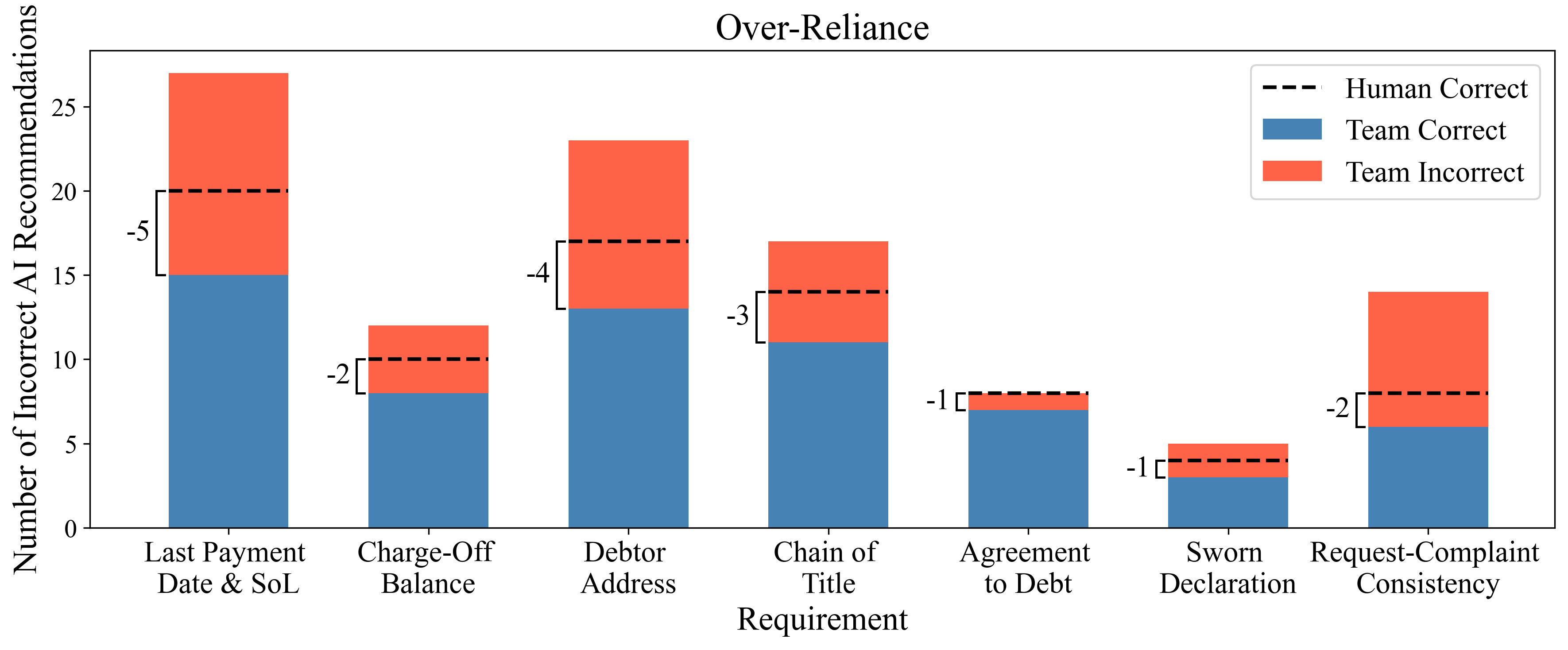}
  \caption{Comparison of the number of incorrect AI recommendations that the assisted humans corrected (blue) to the number of these cases that were correctly reviewed by unassisted humans (dashed line). While the differences are small, there is a general trend of users over-relying on incorrect AI recommendations.}
  \label{fig:overreliance}
\end{figure*}

We observe that users may rely on some incorrect AI recommendations, but do catch the majority of AI mistakes. Given the set of all incorrect AI recommendations for a requirement, we quantify over-reliance by examining the number of cases where assisted humans relied on the incorrect recommendations and unassisted humans otherwise reviewed the requirement correctly (Figure \ref{fig:overreliance}). Though over-reliance affects few cases per requirement, its presence across requirements suggests it remains difficult to fully mitigate.   

\begin{figure*}[h]
  \centering
  \includegraphics[width=1\linewidth]{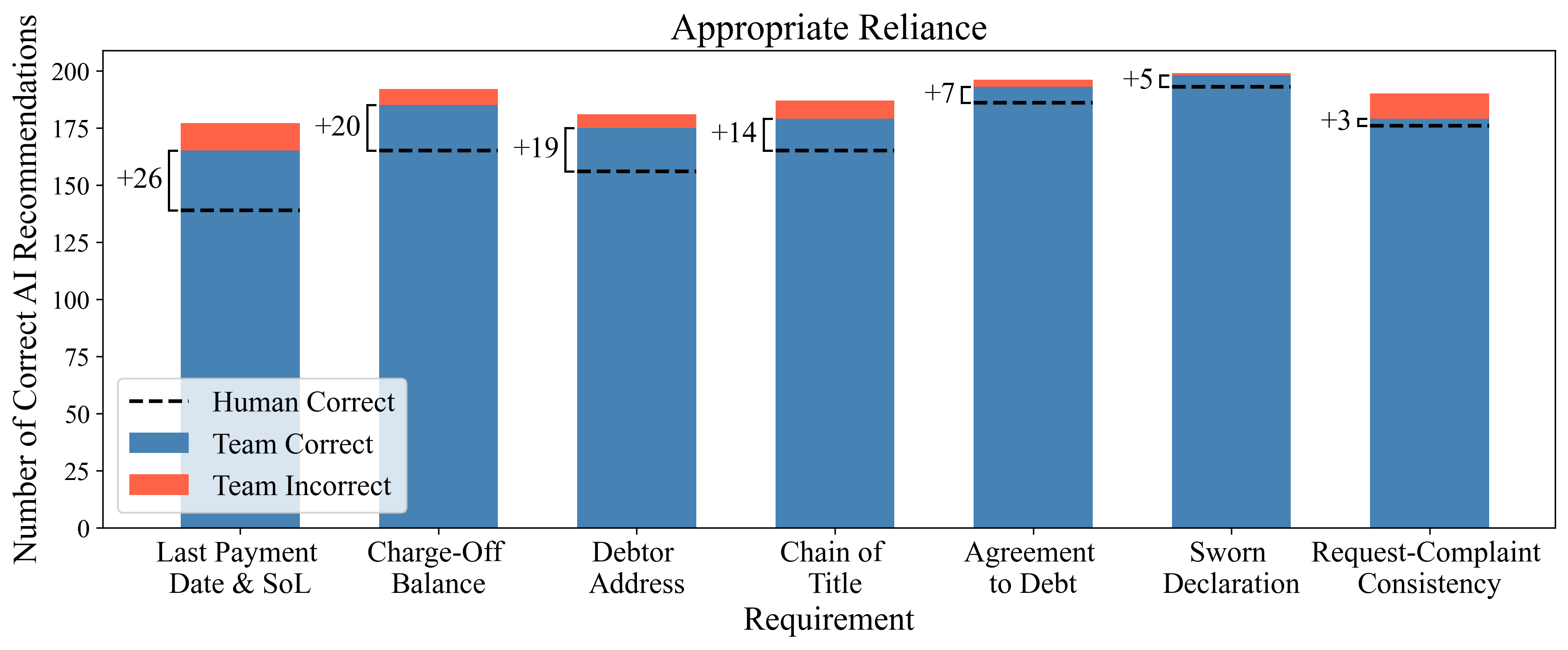}
  \caption{Comparison of the number of correct AI recommendations that the assisted humans agreed with (blue) to the number of these cases that were incorrectly reviewed by unassisted humans (dashed line). Assisted users generally align with correct AI recommendations and through doing so, achieve accuracy gains unrealized by their unassisted counterparts.}
  \label{fig:appropriate_reliance}
\end{figure*}

When it comes to users appropriately relying on correct AI recommendations, we find that assisted users largely, although not perfectly, leverage accurate recommendations. Out of all correct AI recommendations for a requirement, we measure appropriate reliance as the number of cases where assisted humans agree with the correct recommendation and unassisted humans otherwise reviewed the requirement incorrectly (Figure \ref{fig:appropriate_reliance}). By generally adopting correct AI recommendations, assisted users achieve accuracy gains that unassisted users do not. Importantly, the accuracy gains from appropriate reliance counteract the smaller losses from over-reliance.

\section{Fairness Evaluation}

We examine differences in Default Assistant accuracy across the race and gender of defendants and find that they are minimal. Using regex, we find and replace original defendant names with common, racially distinct names selected by \citet{yin_2024} across all pages of files for a case. We focus on defendants because, with rare exceptions, the debt buyer plaintiffs tend to be corporations and not individuals. The augmented defendant names fall into eight categories across perceived race (Asian, Black, Hispanic, and White) and gender (Female and Male). We evaluate each of 155 cases eight times with names randomly selected from the eight categories. 

\begin{table}[h!]
\centering
\caption{Paired differences in average requirement accuracy, aggregated over all permutations of demographic pairs per case (n=155) and tested for equivalence using two one-sided tests (TOST). All comparisons are equivalent within [-0.02, 0.02] with p $<$ 0.05 under TOST, after Benjamini-Hochberg correction.}
\label{tab:fairness_evaluation}
\begin{tabular}{lccc}
\toprule
Group 1 & Group 2 & Avg. Paired Diff. & p-value \\
\midrule
Female & Male & 7.6e-03 & 1.5e-03 \\
Hispanic & Black & -9.2e-04 & 1.0e-02 \\
Asian & White & 1.4e-03 & 1.5e-02 \\
White & Black & 3.7e-03 & 2.4e-02 \\
Hispanic & White & -4.6e-03 & 3.2e-02 \\
Asian & Black & 5.1e-03 & 3.9e-02 \\
Asian & Hispanic & 6.0e-03 & 4.7e-02 \\
\bottomrule
\end{tabular}
\end{table}

Table \ref{tab:fairness_evaluation} shows that observed paired differences are minimal and do not exceed 0.0076. Moreover, per-case paired differences in average requirement accuracy across demographics are bounded within $[-0.02, 0.02]$ with $p < 0.05$ under two one-sided tests. The fact-oriented nature of the requirements evaluated by the Default Assistant—especially given the grounding of recommendations in cited quotes and tables—likely reduces room for bias, in contrast to settings that solicit subjective, open-ended opinions from LLMs. Nonetheless, exploring whether attributes beyond names—such as addresses and transactional patterns—introduce bias into Default Assistant recommendations remains important future work.

\section{Discussion}

% Improvements in accuracy
Our human study demonstrates that the Default Assistant can help users review requirements more accurately and efficiently. While the Default Assistant reduces the time taken by participants to review requirements across the board, the strongest boosts in accuracy are for the \textit{Last Payment Date \& SoL}, \textit{Charge-Off Balance}, and \textit{Debtor Address} requirements where the Default Assistant baseline strongly out-performs the human baseline (Table \ref{tab:complementary_performance}). 

The unassisted participants in our study have a high true rejection rate (Table \ref{tab:major_minor_performance}), compared to the court. The court has historically rejected cases very sparingly. Out of the random sample of 92 cases that requested default judgment, the court did not reject any cases for the requirements reviewed in our audit, despite our finding 6 cases with \textit{Major - Reject} defects. It is possible that the inclination of our unassisted participants to reject cases limited observable improvement in true rejection rate that may otherwise occur with the Default Assistant in the court setting. Nonetheless, we observe that participants aided by the Default Assistant correctly rejected more instances of \textit{Major - Reject} and \textit{Major - Reduce} defects than unassisted participants. Notably, these gains came without substantially lowering true acceptance rates, suggesting the assistant reined in bias rather than inducing over-rejection. In fact, Default Assistant use increased both the true rejection and true acceptance rates for the \textit{Major - Reject} category.  

% Improvements in timing
The \textit{Last Payment Date \& SoL}, \textit{Debtor Address}, \textit{Charge-Off Balance}, and \textit{Sworn Declaration} requirements see the greatest improvements in timing (Figure \ref{tab:team_vs_human_results}). We note that these four requirements all require a user to comb through multiple pages of evidence for specific facts, unlike other requirements which either require referencing a narrowly-scoped location in the case materials (\textit{Agreement to Debt}) or substantial additional reasoning (\textit{Chain of Title}). Future work seeking performance gains from AI assistants may benefit from considering tasks where humans can verify provided answers faster than providing an answer in the first place. 

Our finding that the Default Assistant benefited users in tasks requiring extensive document review might also signal the potential for impact in other contexts. For example, court staff reviewing eviction default judgments must examine leases, rent ledgers, and other exhibits to verify alleged damages and rent owed—a time-consuming process. We hypothesize that an AI assistant that provides recommendations with precise citations might offer significant advantages to diligent court staff.

% Complementary performance and reliance
Interestingly, humans working with the Default Assistant consistently achieve higher accuracy gains than the assistant alone, compared to the human baseline (Table \ref{tab:complementary_performance}). While not shown to be a statistically significant effect, this observation suggests that the users do not rely entirely on the Default Assistant. Indeed, while over-reliance is present, users still catch most AI mistakes. Users and the assistant may make different errors, allowing each to correct the other. Understanding these complementary strengths can guide the design of systems where human-AI teams outperform their individual counterparts.

\section{Limitations \& Future Work}
\label{sec:limitations_and_future_work}

Although we aim to simulate court review as closely as possible, our methodology differs in three key ways. First, our participants were law students who were new to the task of collections case review, despite their general familiarity with legal terminology and case review. In contrast, court staff often have months of experience handling hundreds of cases and receiving judicial feedback. Expertise on the task may impact how effectively users leverage the Default Assistant’s recommendations. Nonetheless, our law student study provides an empirical proof-of-concept that justifies the future investment of SCLAC in a more ecologically valid evaluation.

Second, our unassisted participants presented a strong baseline, rejecting more meritless cases than historical Court judgments. Their performance likely benefited from extra resources: the study solicited decisions through an explicit checklist of requirements—unlike standardized Court protocol—which may itself boost accuracy \citep{haynes2009surgical}. Also, participants were given 10 minutes per case, whereas real-world time constraints clock in at an average of four minutes per case. A shorter time limit may prevent unassisted users from thoroughly reviewing cases, potentially leading AI-assisted users to reject more case defects than users without AI. However, such a reduction in the allotted time per case may also reduce observed relative changes in time efficiency. On the extreme, if users with and without AI assistance run up against the time limit, then there would be no relative efficiency gain.  

Third, participants used the Default Assistant over the course of a limited data collection period. Long-term usage of the Default Assistant over weeks and months may lead to failure modes such as over-reliance or under-reliance. User behaviors and downstream performance may also change as users calibrate their judgment of the strengths and weaknesses of the Default Assistant over longer time periods than a point-in-time study.

Another limitation of our work is that of technological lock-in, where software entrenches institutional practices and can become a barrier to change. We design a modular Default Assistant where statutory requirements may be added, edited, and removed with minimal dependency on other requirements. However, by virtue of encoding institutional knowledge into a software system, we are inevitably increasing the work and skill to evolve that knowledge. This limitation remains an open problem and will require institutional planning and resource allocation to comprehensively address it.   

The Court is continuing to refine its internal review criteria, motivating further development on the Default Assistant to accommodate the new requirements and improve overall performance. As the next step in our staged evaluation, we are planning a year-long study with SCLAC to evaluate the long-term and downstream effects of incorporating the Default Assistant into debt collection default judgment review.

\section{Conclusion}

Our audit of 188 California debt collection cases with default judgments found that 32\% contained defects requiring amendment, 10\% contained inconsistencies requiring reduced judgments, and 4\% contained major defects precluding judgment, highlighting opportunities to reduce erroneous judgments. We built a Default Assistant to support accurate and timely review by courthouse staff, complete with citations to quotes and tables within case files. In a controlled study simulating the court, we studied human-Default Assistant collaboration in terms of accuracy and efficiency. We found that providing reviewers with Default Assistant recommendations citing case materials increased average requirement accuracy by 6.0\% and reduced review time by over 25\% in our study setting. The assistant’s recommendations showed similar accuracy across cases, regardless of defendant names signaling different racial or gender identities.  

While the results from our simulated court setting may not precisely translate to real court settings, our findings establish a proof of concept: the Default Assistant can help reviewers identify case defects more accurately and efficiently. By providing AI recommendations grounded in citations to original case files, we equip reviewers to leverage AI while remaining informed decision-makers. This evidence licenses the next step beyond a simulation to a field study at SCLAC, where the assistant may ultimately help the court reduce unjust wage garnishments, suggest amendments, and more consistently sanction plaintiffs acting in bad faith. Judges are already debating debt collection review procedures, and tools like the Default Assistant may enable more thorough review, potentially shaping future protocols. More broadly, our findings point to AI’s potential to assist with repetitive manual workloads in other courthouse settings.

\section{Acknowledgments}

 We are very grateful to the Superior Court of Los Angeles County judges, research attorneys, staff, and leadership for their continued collaboration, feedback, and openness to innovation. We are deeply grateful to Parker Grove, Victor Wu, and Brian Xu for their invaluable data annotation work, and the many other individuals who have contributed to data annotation throughout the collaboration. We also thank Margaret Hagan for her indispensable help during the study, our participants, and all members of the Rhode Center for their support. Thank you to the participants of JURIX AIDA2J 2025, attendees of ACM CS \& Law 2026, Guestrin and Hashimoto lab members, Myra Cheng, Vishakh Padmakumar, and the SLS law faculty for impactful feedback. 
 
\bibliography{references}

\appendix
\section{Appendix}

\begin{figure*}[h!]
  \centering
  \includegraphics[width=\textwidth]{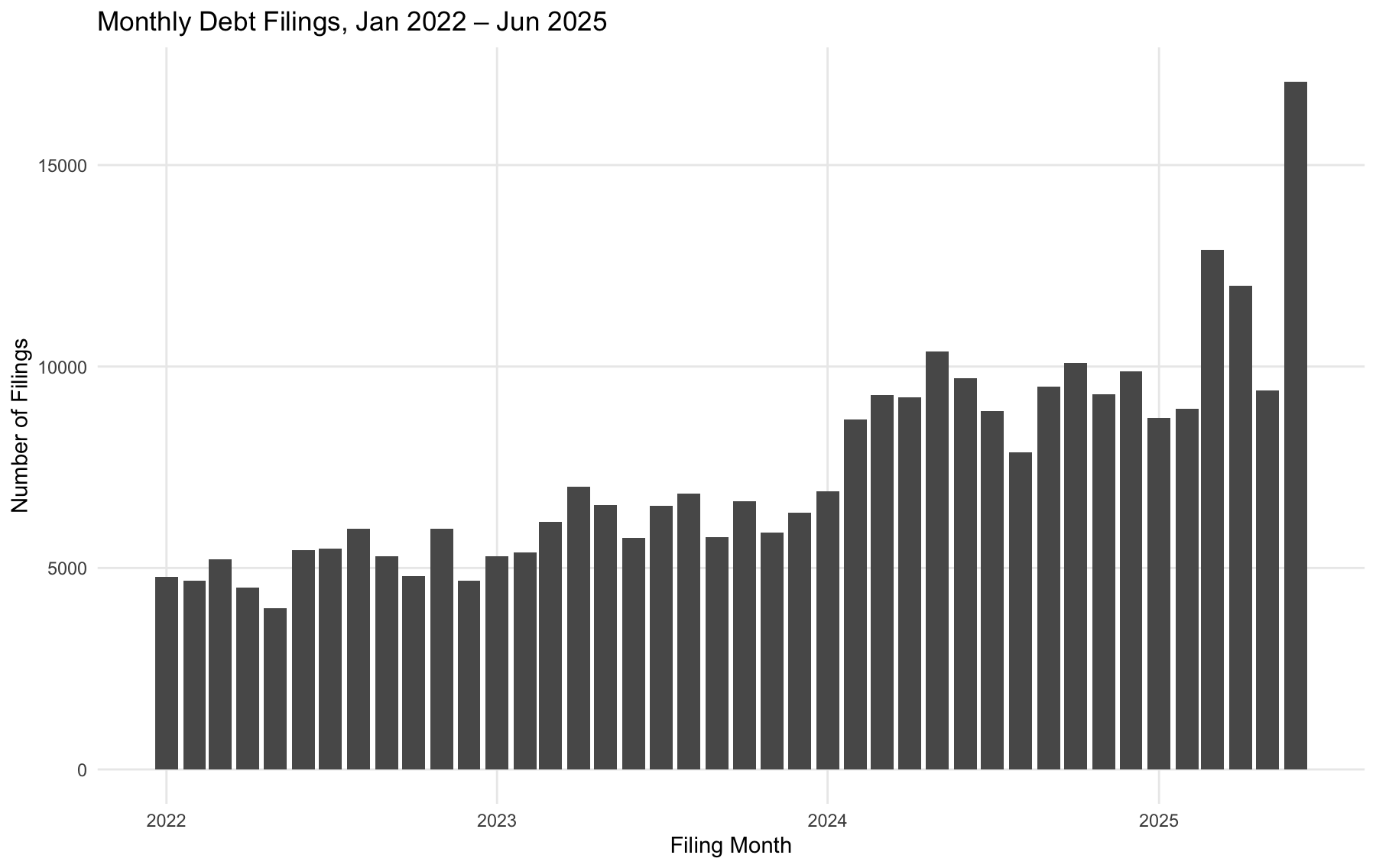}
  \caption{From case management system data, the number of collections cases filed at the Superior Court of Los Angeles has increased since the start of 2022.}
  \label{fig:collection_filings}
\end{figure*}

\begin{figure*}
  \centering
  \includegraphics[width=\textwidth]{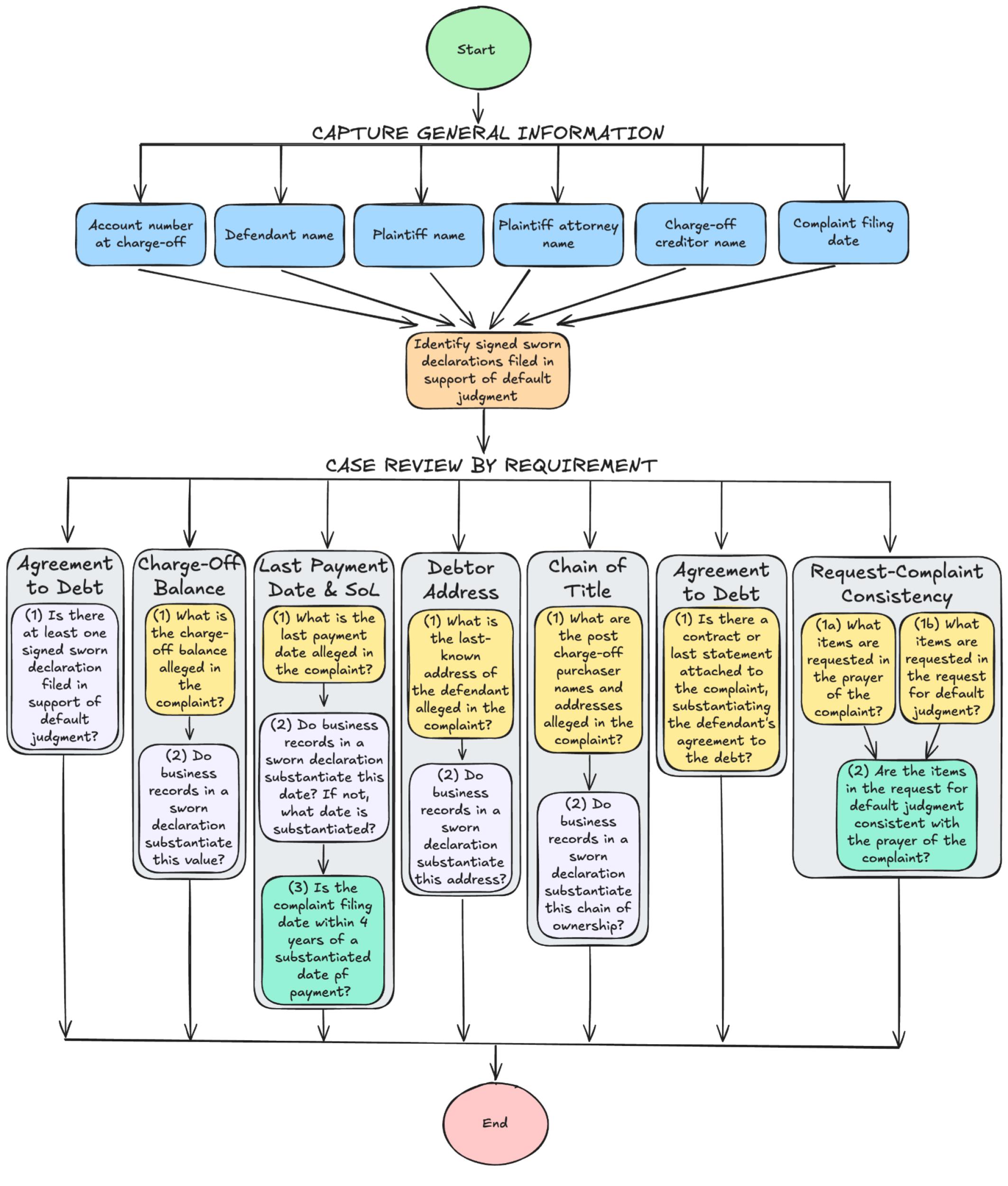}
  \caption{The Default Assistant gathers general case information before reviewing the case for each of the seven requirements. Each box in this graph, aside from ``Start'' and ``End'', retrieve relevant case documents, use information identified by previous nodes in the graph, and generate recommendations with citations for each requirement.}
  \label{fig:default_assistant_graph}
\end{figure*}

\begin{figure*}[ht]
    \centering
    \includegraphics[width=\textwidth]{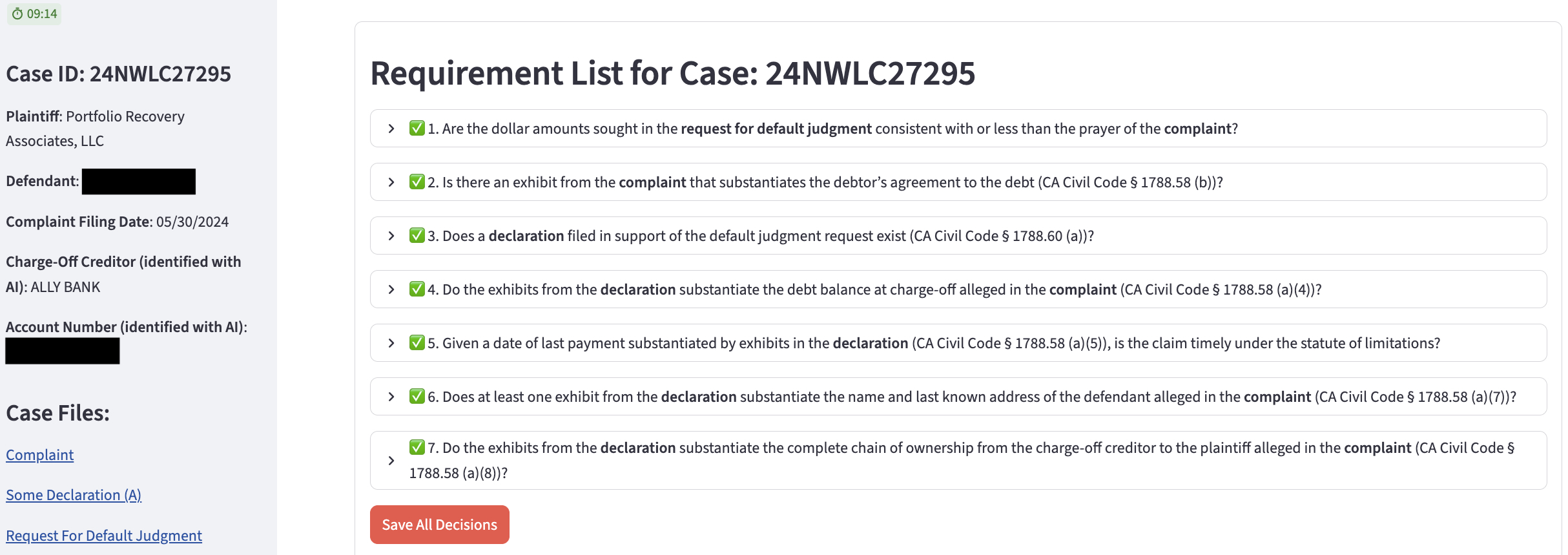}
    \caption{Participants review each case using general case information, case files, and the sub-requirements under the drop-down tabs for each requirement. Only participants assigned to a branch with the Default Assistant are shown ``Charge-Off Creditor (Identified with AI)'' and ``Account Number (Identified with AI)'', in the sidebar. All other case information is pulled from the case management system. Redactions by the authors.}
    \label{fig:ui_req}
\end{figure*}

\begin{figure*}[ht]
    \centering
    \includegraphics[width=\textwidth]{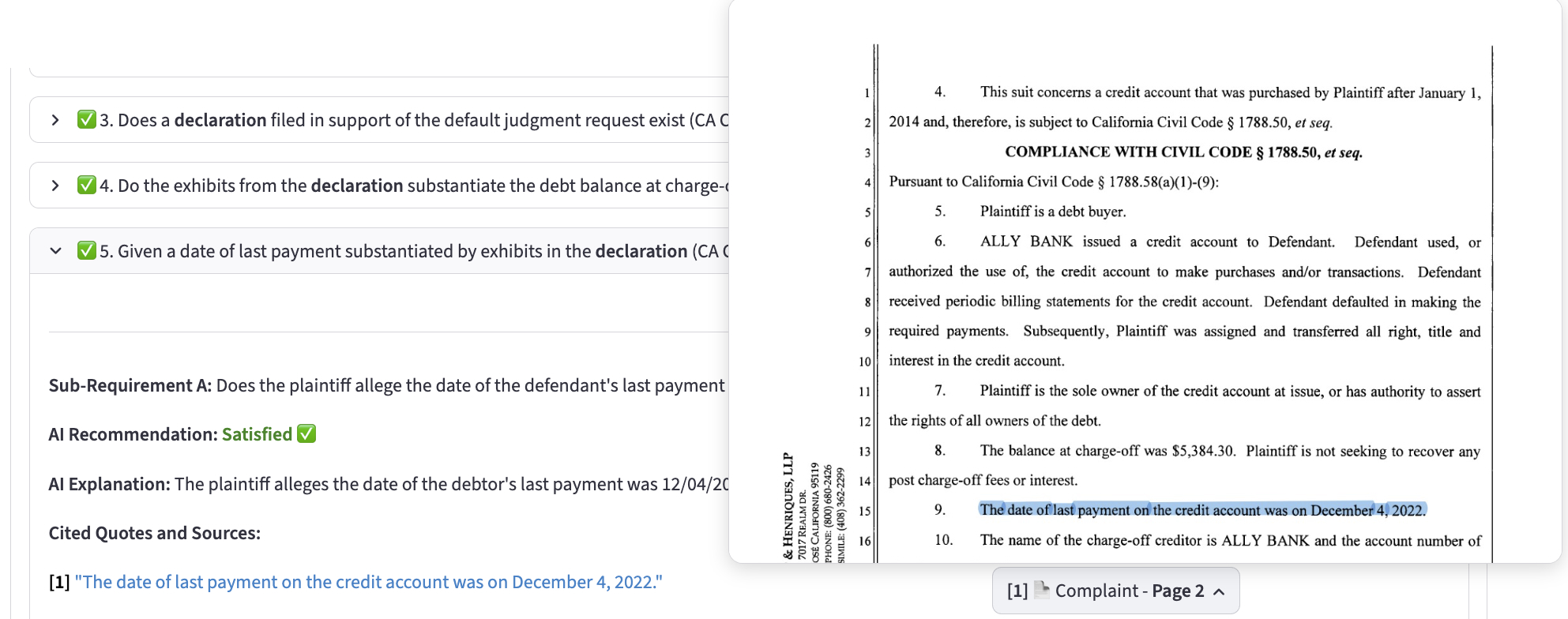}
    \caption{Each citation in the ``Cited Quotes and Sources'' section includes a pop-over that links to the original case-file PDF page with the cited information, highlighted in blue when available. The screenshot shows the page for citation [1] of ``Sub-Requirement A.'' Redactions by the authors.}
    \label{fig:ui_subreq_expanded}
\end{figure*}

\end{document}